\documentclass[referee]{aa} 
\usepackage{graphics}
\usepackage{psfig}
\begin{document}
\def\teff{$T\rm_{eff }$}
\def\lambo{$\lambda$ Boo }
\def\kms {$\mathrm{km\, s^{-1}}$}

\title{ HD 174005: another binary classified as \lambo 
\thanks{Based on
observations collected at the European Southern Observatory (ESO), 
La Silla (Chile) in the framework of the Key Programme 5-004-43K
 }}
   \subtitle{}
\author{ Rosanna \,Faraggiana \inst{1}
\and Mich\`ele Gerbaldi \inst{2,3}
\and Piercarlo Bonifacio\inst{4}
          }

  \offprints{R. Faraggiana}

\institute{
Dipartimento di Astronomia, Universit\`a degli Studi di Trieste,
Via G.B.Tiepolo 11, I-34131 Trieste, Italy \\
email: faraggiana@ts.astro.it 
\and 
Institut d'Astrophysique, 98 bis Bd Arago F-75014 Paris, France \\
email: gerbaldi@iap.fr
\and
Universit\'e de Paris Sud XI
\and
Istituto Nazionale per l'Astrofisica -- 
Osservatorio Astronomico di Trieste, 
Via G.B.Tiepolo 11, I-34131 Trieste, Italy \\
email: bonifaci@ts.astro.it
}

\mail{faraggiana@ts.astro.it}
\date{Received .../Accepted...}
\authorrunning{R. Faraggiana et al}
\titlerunning{HD 174005: another binary classified \lambo}

\date{Received ... / Accepted ...}
\abstract{ 
We demonstrate that
HD 174005, a star recently classified as belonging to the \lambo group, is 
in reality a double lined spectroscopic binary; at some phases, the observed 
composite spectrum may be similar to that of a single star with weak 
metal lines.
}
\maketitle{}

\section{Introduction}

In a recent paper Gray et al (2001) give a precise spectral 
classification of 372 late-A to early-G stars and announce the discovery 
of a new \lambo star:
HD 174005. The spectrum is described as that of a star with the temperature
of an A7-type star but with the K line and metallic line strengths of 
an A2-type star. 

The spectra of the A-F type stars classified as \lambo are 
characterized by weak lines of most metallic species (with the possible 
exception of C,N,O and S); the interpretation of the phenomenon responsible
for these  anomalous atmospheric 
abundances is still controversial in spite
of the large efforts made  to 
define the spectral peculiarities common in these stars
and to interpret the position  of these stars in the HR diagram in 
terms of their evolutionary stage.

The non homogeneous properties of \lambo stars have been confirmed by the 
analyses extended  to spectral ranges other than the classical
optical region. For example the  IRAS data made it possible to detect the 
presence of an IR excess, interpreted as the signature of the presence of 
circumstellar  matter
around some of these stars. 
 A broad absorption feature centred  at 1600 \AA ~ 
has been detected  in some \lambo stars by studying
their IUE spectra; this feature, due to a Lyman ${\alpha}$ satellite, is 
present in white dwarfs and FHB stars too and its detectability in 
\lambo stars
is made possible, in the current opinion, by the lower opacity sources
of \lambo stars compared to "normal" A-type stars.
These are the main, but not the only, heterogeneous properties of this 
class of stars. 

A detailed inspection of the global situation prompted us to 
inspect an alternative explanation of the so called \lambo phenomenon, that of
the duplicity of these stars and we found several observational facts
to support this new hypothesis as described in Faraggiana \& Bonifacio (1999)
and in Faraggiana et al (2001). In fact a composite spectrum, not detected 
as such, can be easily confused with that of a single metal deficient 
star. Several catalogues which provide \lambo classifications are in fact 
contaminated by unrecognized binaries and the abundance analyses of such 
objects, treated as single sources, are meaningless. The purpose of our 
study is to detect the binary stars considered up to now 
as single peculiar objects.

HD 174005 represents a good example to confirm our hypothesis.
We discuss in the next sections the photometric and spectroscopic 
characteristics of this object.

\section{Observations}

\subsection {Optical data}

\subsubsection {Spectroscopy}

We observed HD 174005 twice in the course of an ESO Key Program on
June 9, 1992 and September 7, 1993. The observations have been made at the ESO
1.5m telescope equipped with the Echelec spectrograph; the 
spectra covered 300 
\AA~ centred on H$_{\gamma}$ line with a resolution of about 28000; 
the chosen 
aperture of 320 $\mu$m corresponds to 1\farcs{52} on the sky. 
Details on the original frames and on their reduction and calibration are 
given in Gerbaldi et al (1999).
These spectra have been been already analysed for the measure of stellar
radial velocity by Grenier et al (1999) and the duplicity of this star 
was then detected.

The two spectra have been 
obtained at significantly different phases; the 
duplicity is obvious in the first one, taken in 1992, while
it is not so obvious in the 1993  observation (see Fig. 1).
A large difference in the metal line profiles is easily noticed 
when these two spectra are compared to each other and can be better 
quantified by comparing the observations with theoretical predictions. 

We looked for more observations from the bibliography. We were able to 
find two contradictory stellar classifications: the recent one by Gray et 
al. (2001) referring to a metal weak object and the one by Abt (1988) 
who classifies HD 174005 as an Am star. 

The spectra by Gray et al (2001) on which the classification of HD 174005
as a classic \lambo star is based, were obtained with  
a resolution of 1.8 \AA~ and cover
the wavelength region from 3800 to 4600 \AA; the number of spectra
collected by these authors is not specified. 
The spectra by Abt (1988) have a resolution of 1.0 \AA~. In a previous paper,
Abt (1985) classified this star as an A2 II, with spectra of 2.6 \AA~ 
resolution.

According to our interpretation, these different classifications are 
likely to be the consequence of the spectroscopic variability of this object.

\begin{figure}
\psfig{figure=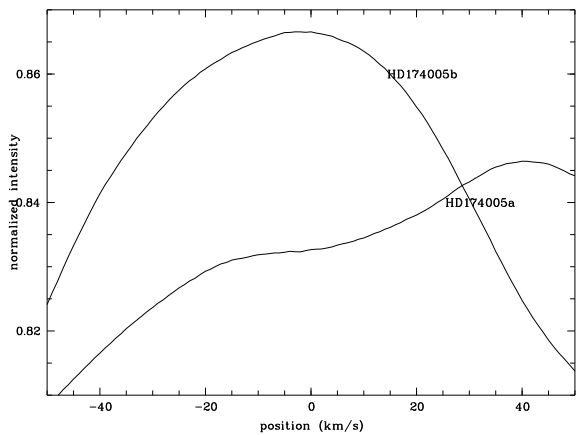,width=8.8cm,angle=-90,clip=true}
\caption{ The cross correlation curves for the two observed spectra 
(HD 174005a taken on
June 9, 1992 and HD 174005b taken on Sept. 7, 1993); the 
template is the synthetic spectrum computed by using the model described
in the text.}
\label{fig1}
\end{figure}

\begin{figure}
\psfig{figure=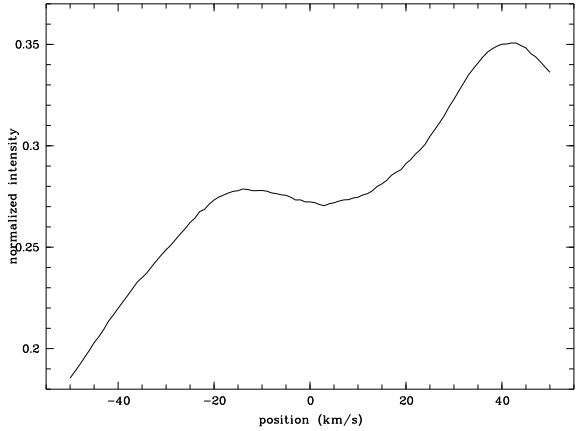,width=8.8cm,angle=-90,clip=true}
\caption{ The cross correlation curve for only the metal lines of the 1992 
spectrum obtained by using as template the synthetic spectrum computed 
with the model with  the Geneva atmospheric parameters.
}
\label{fig2}
\end{figure}

\begin{figure*}
\psfig{figure=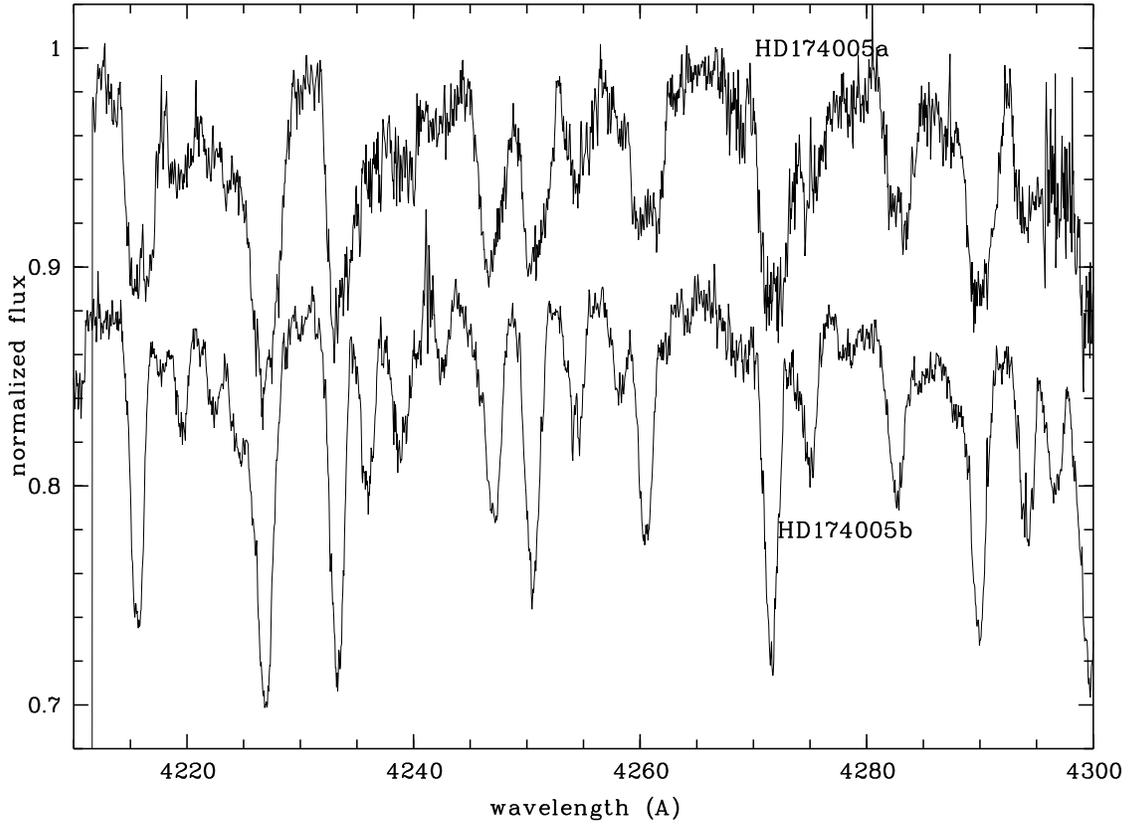,width=17cm,angle=-90,clip=true}
\caption{ The comparison of the two observed spectra in the region 4200-4300 
\AA ; HD174005a and HD 174005b refer to spectra taken on June 9, 1992 and 
on Sept. 7, 1993 respectively.}
\label{fig3}
\end{figure*}

\begin{figure*}
\psfig{figure=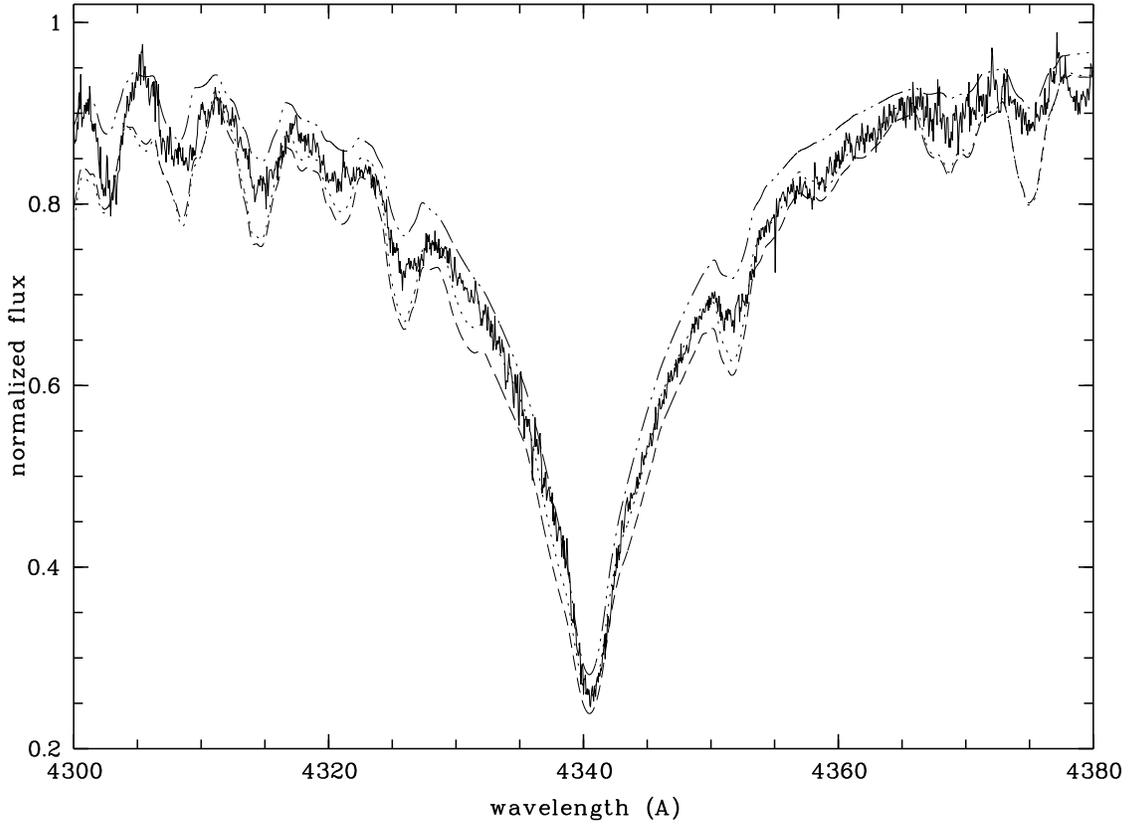,width=17cm,angle=-90,clip=true}
\caption{ The H$_{\gamma}$ profile of HD 174005 spectrum observed in 1992
(solid line) compared with the spectra from the models: \teff=7673 K, 
log g=3.98, $v\sin i =100 $km s$^{-1}$ and  solar abundances (dotted line), 
the abundances reduced
by 10 (dash-dot-dot line) and \teff=7910 K, log g=3.79 and the same broadening
(dashed line). 
 }
\label{fig4}
\end{figure*}

\begin{figure*}
\psfig{figure=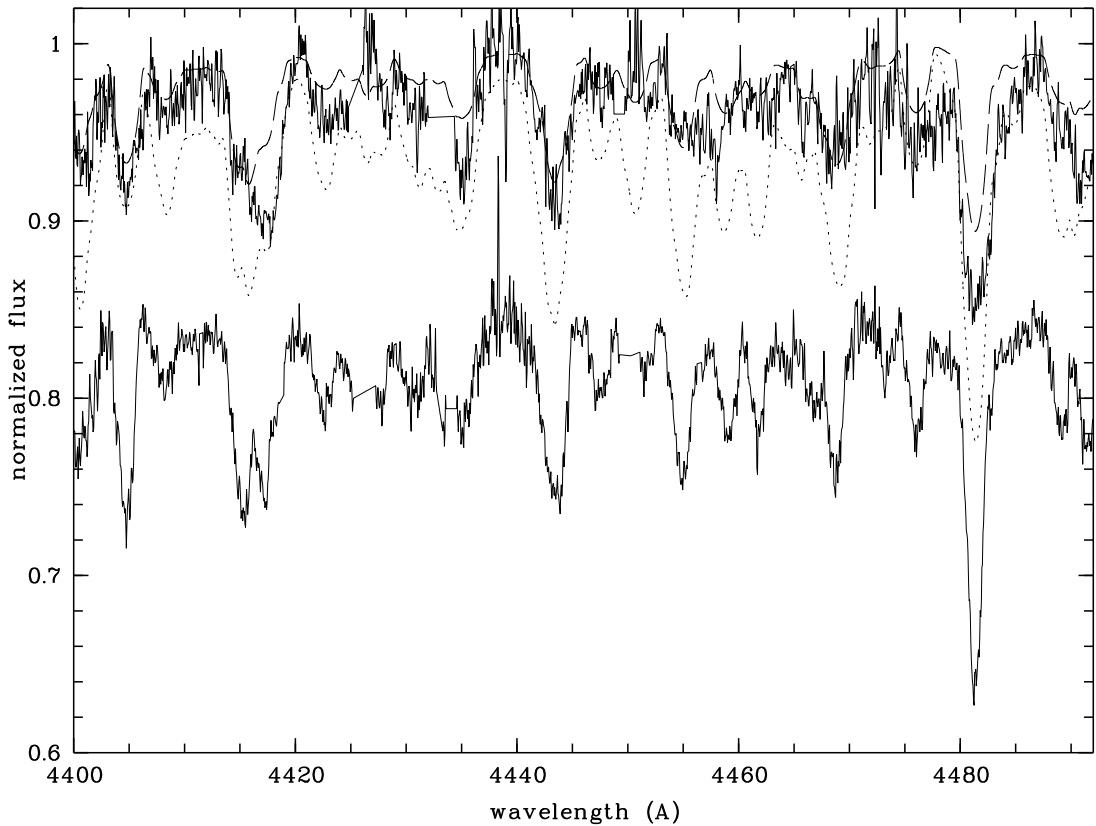,width=17cm,angle=-90,clip=true}
\caption{ The poor fit of the metal lines with the computed spectra computed
assuming the model with \teff=7673 K, log g=3.98 and solar abundances (dotted 
line) and metal abundances reduced by a factor of 10 (dashed line); we note
in particular that the Mg II 4481  doublet in the spectrum taken in 1992 does 
not have a rotationally broadened profile and the feature at 4415 \AA~ is
not reproduced by the computations. The straight lines in the observed 
spectra correspond to bad CCD columns.
 }
\label{fig5}
\end{figure*}

\begin{figure}
\psfig{figure=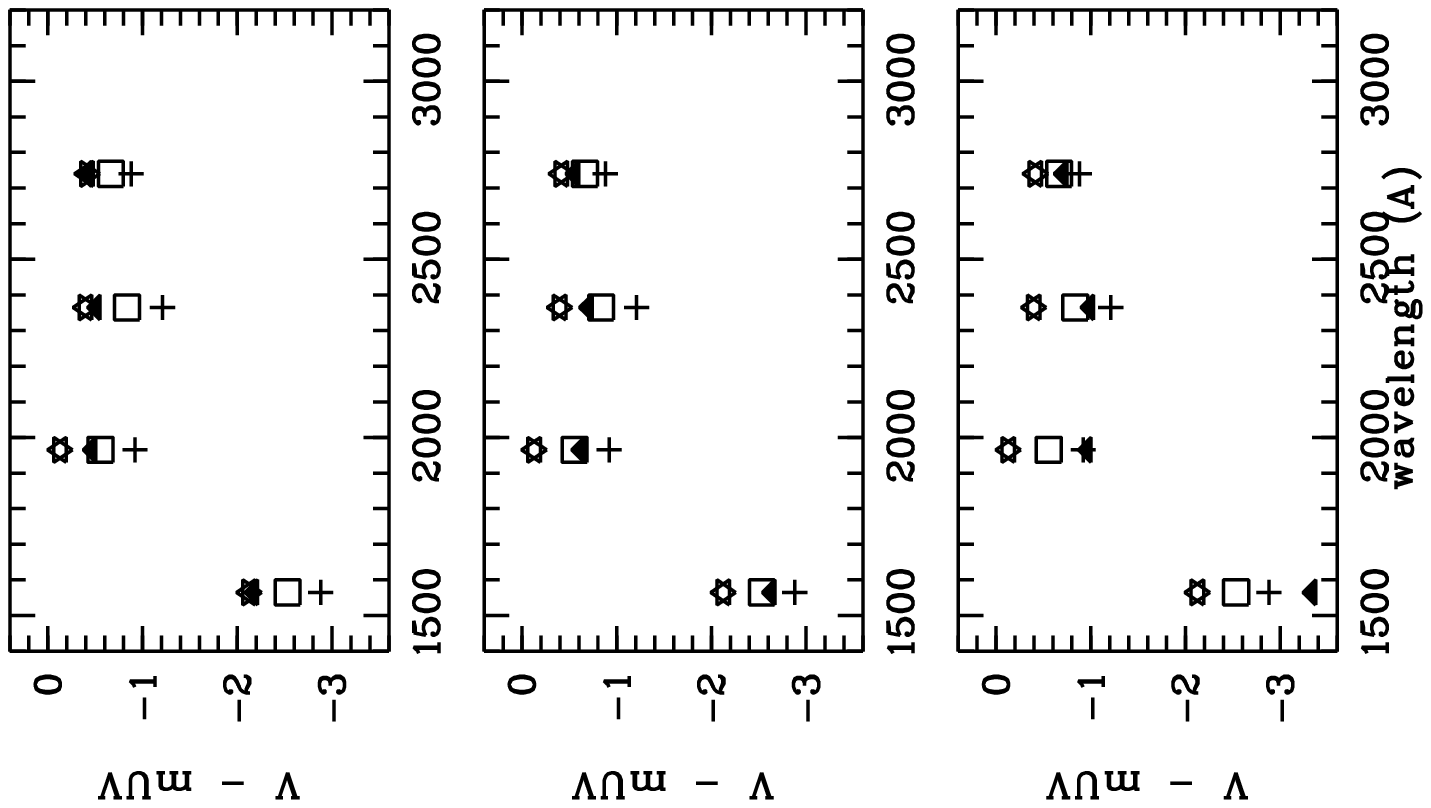,width=8.8cm,angle=-90}
\caption{The observed and dereddened UV colour indices (V-m$_{UV}$)$_o$
compared with those expected for a star with the parameters \teff = 7750 K,
log g =4.0 and [Fe/H] = 0. (bottom), -0.5 (middle) and -1. (top) labelled
with filled triangles; the symbols for the observed indices
are +: undereddened values; square: E(B-V)=0.069; star: E(B-V)=0.152.
  }
\label{fig6}
\end{figure}

\subsubsection {Photometry}

The information we could retrieve from the literature on this star are 
quite scanty; the 
star has been observed photometrically only by Oblak E. (1978) in 
uvby$\beta$ (two observations) and by Hall \& Mallama (1974) and by Eggen
(1968) once and twice respectively in UBV, so that a
possible photometric variability could have passed unobserved.

The photometric observations by the Tycho experiment on board of the Hipparcos
space experiment span over 3 years; no variability was detected
from the 83 observed values, according to the Hipparcos Main Catalogue (ESA
1997); in fact the star is labelled "C"= constant. 
HD 174005 (= BD$-06^\circ 4913$ = TYC 5126-2404)
             has  a nearby visual companion HD 174005B 
            (= BD$-06^\circ 4912$ = TYC 5126-2381), the similar proper
            motion of this pair, as given in the Tycho-2 Catalogue 
            (Hog et al. 2000) suggests that they are physically linked.
            However since the parallax has not been measured
            for HD 174005B a doubt remains on
           the physical association of these two objects.

From the uvby$\beta$ colour indices we derived the interstellar absorption 
by using the Moon (1985) code. 
The resulting E(b-y)=0.050 is slightly 
larger than that obtained for HD 174005 B 
by the same procedure: E(b-y)=0.030.
By assuming the hypothesis that the two stars are physically associated,
i.e. have the same distance, the latter E(b-y)
value is coherent with the results of the analysis by Vergely et al (1998)
of the interstellar extinction in the solar neighbourhood.

\subsection {UV data}

HD 174005 has not been observed by IUE, but only by the S2/68 
on board of the ESRO
TD1 satellite. The measured fluxes at 2740, 2365, 1965 and 1565 \AA~ allow to
determine the stellar reddening on the basis of the UV colours only, as
discussed in the Catalogue of stellar UV fluxes (Thompson et al, 1978). The 
approximate relation for A-type stars:
$$\mathrm 
E(B-V) = 0.43(m_{1565}-m_{2765})-0.46(m_{1565}-m_{2365})+0.06
$$
has been used to derive an independent reddening value.

The extinction E(B-V)=0.152 so obtained is much higher than
that derived from the visual uvby$\beta$ photometry, E(B-V)=0.069.
We shall discuss these UV data in the next section with theoretical
computations of the UV fluxes.

\section {Stellar parameters and spectral synthesis}

\subsection {Atmospheric parameters}

The atmospheric parameters derived from the uvby$\beta$ colour indices
by using the Moon and Dworetsky (1985) code and adopting the previously 
derived reddening, E(b-y)=0.050, are  \teff=7910 K and log=3.79. 

We have also used the Geneva photometry indices to compute \teff~
and log g, by adopting the same reddening value and by using the K\"unzli
et al (1997) calibration; in this case the parameters are:
\teff= 7673 $\pm$ 65 K, log g =   3.98 $\pm$ 0.06 and 
[M/H]  =  -0.77 $\pm$     0.24.

\subsection {HR diagram}

These data allow us to put the star in the HR diagram.
For that purpose we used the evolutionary tracks computed
by Schaller et al (1992) and the isochrones by Meynet et al (1993)
for the solar chemical composition.

The absolute magnitude corrected for the reddening is:  M$_V$=0.27;
by applying the bolometric correction,  taken from the Bessell et al 
(1996) tables, the bolometric magnitude results M$_{bol}$=0.22.
By adopting M$_{bol\odot}$=4.75 the luminosity is derived:
log (L/L$_{\odot}$)=1.81.
According to the position of the star in the HR diagram 
the value
of log g is of about 3.6, slightly  lower than the 
two photometrically derived values 3.79 and 3.98. 

\subsection {Computed spectra}

We have computed a synthetic spectrum 
by using the Kurucz model with \teff=8000 K, log g=4.0 and solar abundances 
to investigate the correlation with the observations.
The normalized cross correlation indices between the two 
observed spectra and the theoretical one are displayed in Fig. \ref{fig1}.
The two peaks obtained for the 1992 spectrum are evident, while  
only a broad and asymmetric cross correlation curve results from the 1993 
data; however we note the low value of the normalized correlation peak 
even at this date.

A further correlation (Fig. \ref{fig2}) 
of only the metallic lines with as template 
the spectrum
computed by using the model with the Geneva atmospheric parameters shows even
better the two peaks for the 1992 spectrum and allows to measure the 
differential radial velocity of the two components of this spectroscopic
binary system at this phase: $\Delta$RV= 52 km s$^{-1}$.

The large variations in the line profiles of the metal lines are shown 
in Fig. \ref{fig3}, 
where a sample of the 2 spectra, corrected for the radial velocity 
values of the brighter component are plotted.
The star appears as a clear SB2 on the 1992 spectrum; we note in particular
the well resolved Sr II 4215 line, the double peaks of the Fe I 4271 and
of the Cr I-Ti II 4290 blends and the general broader profiles in this
spectrum labelled as HD174005a.

As a further sign of duplicity 
we recall that the observed core of Balmer lines should be deeper than that
in the spectrum computed by using Kurucz SYNTHE code which does not
include NLTE effects. In the contrary, the spectrum taken in 1992 
has the H$_{\gamma}$ core shallower than or as deep as  the computed ones.
In Fig. \ref{fig4} 
this observed profile is compared with those computed by using the 
models with the parameters derived in section 3.1. The best fit of 
H$_{\gamma}$ is obtained
by using the parameters \teff=7673 K, log g=3.98 and solar abundances; however 
this choice does not reproduce the metallic lines as may be seen in Fig.
\ref{fig5}.  

The broad and apparently weak metal lines in the 1992 spectrum are 
given in Fig. \ref{fig5} 
where are compared with two computed spectra (\teff=7673 K,
log g=3.98), one with
solar abundances and the other one with these reduced by 10.; we note in
particular the non-rotationally broadened profile of Mg II 4481 which is 
characterized by a very flat and
square core in the 1992 spectrum. 

Having only two spectra at our disposal and taken at more
than one year of interval, we cannot make any guess on the possible
period of this system.

\subsection{UV photometry}

We have compared the S2/68 TD1 observations,
dereddened by using  E(B-V)=0.069 or 0.152, with
theoretical indices  computed by integrating 
the Kurucz fluxes for several metal-abundances,
over the band-width
of the four S2/68 TD1 channels.

The observed and theoretical colour indices (V$-$m$_{UV})_0$ 
are  shown in Fig. \ref{fig6}.
The flux energy distribution of HD 174005 in the UV does not indicate
a significant metal underabundance unless the unrealistic high reddening
E(B-V)=0.152 is assumed.
The observations  fit  the computations with a moderate
underabundance  between  $-1.0 $ and $-0.5$  dex.
This is coherent with the 
value of $-0.77$ derived from the Geneva photometry.

The coherence of the flux distribution over a large wavelength range
from the optical to the UV reinforces the result obtained from spectroscopy
in the optical:  the two stars have a very similar luminosity and
effective temperature.

\section { Conclusions }

We have shown that  HD 174005 is a 
spectroscopic binary and the observed spectrum is highly variable
with the phase.  
The confusion between the spectrum of a single weak lined star
and the composite spectrum of a binary composed by two similar, but not 
equal sources, is demonstrated. 

A careful check of high dispersion data and the analysis based on more 
than a single spectrum appear to be necessary to obtain reliable
abundance analyses of \lambo candidates; as a first step it is necessary 
to distinguish composite spectra from those of single weak-lined stars and 
this can require a quite subtle analysis of mean-high resolution spectra.
 


\end{document}